\documentclass[reprint,groupedaddress,superscriptaddress,amsmath,amssymb,aps,twocolumn,floatfix, prl,table]{revtex4-2}
\usepackage{graphicx}
\usepackage{dsfont}
\usepackage{rotating}
\usepackage{amsmath}
\usepackage{amsthm}
\usepackage{amssymb} 
\usepackage{array}
\usepackage{dcolumn}
\usepackage{subcaption}
\usepackage{tikz}
\usetikzlibrary{positioning, backgrounds}
\usepackage{appendix}
\definecolor{mygreen}{rgb}{0.2, 0.8, 0.2}
\usepackage[colorlinks=true, linkcolor=blue, citecolor=mygreen, urlcolor=red]{hyperref}
\usepackage{hhline}
\usepackage{listings} 
\usepackage{booktabs} 
\usepackage{bm}
\usepackage{tikz}
\usepackage{multirow}
\usepackage{array}
\usepackage{comment}
\usepackage{colortbl}
\usepackage{comment}
\usepackage[compat=1.1.0]{tikz-feynman}
\usepackage{braket}
\usepackage{simpler-wick}
\usepackage{MnSymbol}
\usepackage{eucal}
\usepackage{epstopdf}
\usepackage{lipsum}
\usepackage{float}
\usepackage{svg}
\usepackage{tikz}
\usetikzlibrary{decorations.pathmorphing, decorations.markings, arrows.meta}
\usepackage{dcolumn}
\usepackage{eqnarray, amsmath}
\usepackage{amsthm}
\usepackage{scalerel}
\usepackage{stackengine,wasysym}
\usepackage{yfonts}
\usepackage{bbm}
\usepackage{mathbbol}
\usepackage{amsfonts}
\usepackage{dsfont}
\usepackage{titlesec}
\usepackage{xr-hyper}
\usetikzlibrary{positioning,calc,decorations.pathreplacing,arrows.meta}
\usepackage{tikz}   
\usepackage{algorithm}
\usepackage{caption}
\usepackage{algpseudocode}
\makeatletter
\def\thmhead@plain#1#2#3{%
  \thmname{#1}\thmnumber{\@ifnotempty{#1}{ }#2}%
  \thmnote{ {\the\thm@notefont(#3)}}}
\let\thmhead\thmhead@plain
\makeatother

\newtheorem{definition}{Definition}
\newtheorem{theorem}{Theorem}

\newcommand{\twocolcaption}[1]{%
  \par\vspace{0.5em}%
  \noindent
  \begin{minipage}[t]{0.94\textwidth}  
    \raggedright
    \small #1
  \end{minipage}
}

\begin{document}
\title{Fast summation of fermionic Feynman diagrams beyond Bravais Lattices and on-site Hubbard interactions}
\author{Boyuan Shi}
\email{boyuanshi0607@gmail.com}
\affiliation{Blackett Laboratory, Imperial College London,
London SW7 2AZ, United Kingdom}

\begin{abstract}
We designed new algorithms for summing bold-line Feynman diagrams in arbitrary channels, where it can be readily modified for bare interaction series as well. When applied to magnetic channel bold-line series with on-site Hubbard interactions, the algorithm achieves competitive performance compared with the state-of-art RPADet. We then generalize it beyond square lattice and on-site Hubbard interactions and achieve better scaling in the number of sites within a unit cell, while there is substantial increase when there are more types of interactions. This work paves the way of diagrammatic Monte Carlo simulations for real materials, holding the premise for a robust replacement of state-of-art simulation tools in the thermodynamical limit.
\end{abstract}
\maketitle
\textit{Introduction and Background}-- Accurately simulating quantum many-body systems can elucidate a broad spectrum of phenomena across materials sciences and chemistry. Diagrammatic Monte Carlo methods are among the few numerical methods that could obviate finite-size extrapolations while also maintaining high accuracy, which holds the promise of explicating intricate long-range quantum correlations and nuanced spectral signatures intrinsic to both exotic and conventional manifestations of condensed matter.

Its development has traversed a protracted and tortuous historical trajectory. It was firstly applied for systems without fermionic sign problems, \textit{e.g.} for polarons \cite{prokofev_svistunov_polaron_2000}, where one performs random walks in the space of Feynman diagrams with judicious choices of local diagrammatic updates.

In the context of problems not immune to the sign problem, the conventional paradigm, which emphasizes meticulous refinement of the proposal function, is eclipsed by the factorial suppression of the average sign concomitant with ascending orders in the diagrammatic expansion. In particular, the worm algorithm for the Hubbard model fails to reach high enough expansion orders, constituting a formidable impediment for subsequent advancements.

Soon it was realized that the sum of all Feynman diagrams with contact Hubbard interactions at a given order can be written as the product of two determinants \cite{contact_interaction_determinant}, and removing disconnected diagrams can be done with simple set recursions \cite{Rossi_CDet}. Subsequently, this method underwent algorithmic refinement, culminating in the contemporary state-of-the-art methodology applicable to this type of systems. Fermionic models in this family are then able to be studied extensively, with landmark findings encompassing predictions of metal-to-insulator transitions at half-filling \cite{MIT_Half_filled_DiagMC}, high-resolution spin and density correlations \cite{2D_Hubbard_Model_CDet_PRR}, origin and the fate of pseudogap under doping \cite{pseudogap_CDet}, \textit{etc}.

Behind the veil of prosperity is the limited range of applicability restricted to the Hubbard model with contact interactions in the Bravais lattices. Moreover, almost all the previous studies that could reach large expansion orders employs weak-coupling expansions. At strong interactions, minute and nuanced variations in the curve of thermodynamical quantities that are completely absent from DiagMC prediction can be reveled by tensor-network methods in small clusters. Assuming that they are not finite-size artifacts, more refinements are required to modify the bare perturbation series to render it obtaining intrinsically non-trivial results.

A natural way out of the dilemma would be to choose series with larger radius of convergence, \textit{e.g.} with re-normalized Green's functions and interactions. However, data points at large expansion orders remain scarce at present, predominantly owing to the substantial escalation in computational expense relative to the bare approach. Also, it was seen in \cite{shi2022one_over_N} that a good series, \text{e.g.} $1/\mathrm{N}_f$ amounts doing powerful re-summation already, while only a few orders are needed to get accurate results given a ``physical" starting point of the expansion. There are essentially an infinite number of such schemes, while judicious changes in graphic structures require more flexible algorithms. In momentum space, we are able to reach high expansion orders without the use of combinatorial algorithms \cite{shi2022one_over_N}. Employing fast summation techniques in momentum space would confer substantial computational advantages and markedly enhanced numerical stability, attributable primarily to the significantly more compact configurations involved. 

All of those motivate the design of new combinatorial platforms that could be flexible and efficient in those complicated scenarios. A solid foundation for further algorithmic developments is the existence of a powerful ``global" sampling method without local diagrammatic updates, which is already fully tested in our previous constructions \cite{shi2022one_over_N}. Our new algorithm is based on the idea of bosonization in field theory, which enables systematic contacts with well-studied algebraic structures like permanent and hafnian. In the following, we will demonstrate the key idea with contact Hubbard interactions in arbitrary lattices, while the generalizations to other configuration are straightforward.   

\textit{Field Theoretical Constructions}-- 
Rather than the $G-\Sigma$ bosonization approach in our previous paper \cite{shi2022one_over_N} in the density-density channel, we perform the functional replacement in the magnetic channel with the action given by
\begin{equation}
\begin{aligned}
&S=\sumint_{\tau,k,l,\sigma}\;\bar{\psi}_{\sigma}(k,\tau)\left[(\partial_{\tau}-\mu)\delta_{kl}+h_{kl}\right]\psi_{\sigma}(l,\tau)\\
&-\frac{1}{2}\sumint_{\tau,k,l}\;V_{\sigma\sigma'}(k,l)G_{\sigma'\sigma}(l,k;\tau,\tau^+)G_{\sigma\sigma'}(k,l;\tau,\tau^{+})\\
&-i\sumint_{\tau,\tau',k,l}\;\Sigma_{\sigma'\sigma}(l,k;\tau',\tau)[G_{\sigma\sigma'}(k,l;\tau,\tau')-\\
&\qquad\qquad\qquad\qquad\qquad\quad\;\;\;\psi_{\sigma}(k,\tau)\bar{\psi}_{\sigma'}(j,\tau')],
\end{aligned}
\end{equation}
where as in \cite{shi2022one_over_N}, we have introduced two bosonic fields $G$ and $\Sigma$. $i$ is the imaginary number unit. The saddle point self-energy is given by $i\Sigma_{*}(1,2)=-V(1,2)G_{*}(1,2)$, where we have introduced the compact index notations with $1\equiv(\tau_{1},i_{1},\sigma_{1},..)$. While in the density-density channel $i\Sigma(1,2)=\sumint_{3}v(1,3)G(3,3)\delta_{1,2}$. We will discuss general mixed channel case later but we find understanding the magnetic channel solely with on-site interactions are heuristic enough to witness the power of the ``bosonized algorithm". 

Perturbation series after bosonization starts with expanding the action around the saddle point to quadratic order, with the four fixed second-order derivatives given by
\begin{equation}
\begin{aligned}
&\frac{\delta^2S}{\delta G(2,1)\delta G(3, 4)}=-V(1,2)\delta_{1,3}\delta_{2,4}\equiv W_{G}(1,2;3,4),\\
&\frac{\delta^2S}{\delta G(2,1)\delta \Sigma(3, 4)}=-i\delta_{2,4}\delta_{1,3}\equiv -i\mathds{I},\\
&\frac{\delta^2S}{\delta \Sigma(2,1)\delta\Sigma(3,4)}=-G_{*}(4,2)G_{*}(1,3)\equiv-W_{\Sigma}(1,2;3,4),
\end{aligned}
\end{equation}
where $G_{*}$ is the saddle-point Green's function. The important correlator to compute perturbation series of $\mathrm{ln}Z$ is $\langle\delta\Sigma(1,2)\delta\Sigma(3,4)\rangle=[(1-W_{G}W_{\Sigma})^{-1}W_{G}](2,1;3,4)$.

The geometric series of $\langle\delta\Sigma(1,2)\delta\Sigma(3,4)\rangle$ now orients in the different way as that in the density-density channel \cite{shi2022one_over_N}
\begin{equation}
\begin{tikzpicture}
[scale=0.75]
    \coordinate (p4) at (0,2);
    \coordinate (p1) at (1.5,2);
    \coordinate (p2) at (1.5,0.5);
    \coordinate (p3) at (0,0.5);
    
    \foreach \i/\pos in {1/{right}, 2/{right}, 3/{left}, 4/{left}} {
        \fill (p\i) circle (2pt);
        \node[\pos] at (p\i) {$\i$};
    }
    
    \draw[loosely dashed] (p4) -- (p1);
    
    \draw[decorate, decoration={snake, amplitude=0.5mm, segment length=3mm}, blue] (p1) -- (p2);
    
    \draw[loosely dashed] (p2) -- (p3);
    
    \node at (2.5,1.25) {$+$};
    
    \coordinate (q4) at (3.5,2);
    \coordinate (q1) at (5,2);
    \coordinate (q2) at (5,0.5);
    \coordinate (q3) at (3.5,0.5);
    
    \foreach \i/\pos in {1/{right}, 2/{right}, 3/{left}, 4/{left}} {
        \fill (q\i) circle (2pt);
        \node[\pos] at (q\i) {$\i$};
    }
    
    \draw[thick, decoration={markings, mark=at position 0.55 with {\arrow{Triangle[black, width=4pt, length=4pt]}}}, postaction={decorate}] (q4) to[out=30, in=150] (q1);
    
    \draw[decorate, decoration={snake, amplitude=0.5mm, segment length=3mm}, blue] (q1) -- (q2);
    
    \draw[thick, decoration={markings, mark=at position 0.55 with {\arrow{Triangle[black, width=4pt, length=4pt]}}}, postaction={decorate}] (q2) to[out=210, in=-30] (q3);
    
    \draw[decorate, decoration={snake, amplitude=0.5mm, segment length=3mm}, blue] (q3) -- (q4);
    
    \node at (6,1.25) {$+$};
    
    \coordinate (r4) at (7,2);      
    \coordinate (r4p) at (7.75,2);  
    \coordinate (r1) at (8.5,2);    
    \coordinate (r3) at (7,0.5);    
    \coordinate (r3p) at (7.75,0.5); 
    \coordinate (r2) at (8.5,0.5);  
    
    \fill (r4) circle (2pt);  \node[left] at (r4) {$4$};
    \fill (r4p) circle (2pt); \node[above] at (r4p) {$4'$};
    \fill (r1) circle (2pt);  \node[right] at (r1) {$1$};
    \fill (r3) circle (2pt);  \node[left] at (r3) {$3$};
    \fill (r3p) circle (2pt); \node[below] at (r3p) {$3'$};
    \fill (r2) circle (2pt);  \node[right] at (r2) {$2$};
    
    \draw[thick, decoration={markings, mark=at position 0.65 with {\arrow{Triangle[black, width=4pt, length=4pt]}}}, postaction={decorate}] 
          (r4) to[out=30, in=150] (r4p);
    
    \draw[thick, decoration={markings, mark=at position 0.65 with {\arrow{Triangle[black, width=4pt, length=4pt]}}}, postaction={decorate}] 
          (r4p) to[out=30, in=150] (r1);
    
    \draw[thick, decoration={markings, mark=at position 0.55 with {\arrow{Triangle[black, width=4pt, length=4pt]}}}, postaction={decorate}] 
          (r2) to[out=210, in=-30] (r3p);
    
    \draw[thick, decoration={markings, mark=at position 0.6 with {\arrow{Triangle[black, width=4pt, length=4pt]}}}, postaction={decorate}] 
          (r3p) to[out=210, in=-30] (r3);
    
    \draw[decorate, decoration={snake, amplitude=0.5mm, segment length=3mm}, blue] (r1) -- (r2);
    
    \draw[decorate, decoration={snake, amplitude=0.5mm, segment length=3mm}, blue] (r3p) -- (r4p);
    
    \draw[decorate, decoration={snake, amplitude=0.5mm, segment length=3mm}, blue] (r3) -- (r4);

    \node at (9.5,1.25) {$+\cdots$};
\end{tikzpicture}
\label{eq:RPA_series}
\end{equation}
If the interactions are purely contract, $V(1,2)=(\delta_{\sigma_{1},\uparrow}\delta_{\sigma_{2},\downarrow}+\delta_{\sigma_{1},\downarrow}\delta_{\sigma_{2},\uparrow})\delta(\tau_{1}-\tau_{2})\delta_{\boldsymbol{R}_{1},\boldsymbol{R}_{2}}U$, then $\langle\delta\Sigma(1,2)\delta\Sigma(3,4)\rangle$ can be decomposed into two parts with 
\begin{equation}
\begin{aligned}
\langle\delta\Sigma(1,2)\delta\Sigma(3,4)\rangle=&[\delta_{\sigma_{1},\downarrow}\delta_{\sigma_{2},\uparrow}\delta_{\sigma_{3},\uparrow}\delta_{\sigma_{4},\downarrow}W_{\uparrow\downarrow}(2,3)+\\
&\;\delta_{\sigma_{1},\uparrow}\delta_{\sigma_{2},\downarrow}\delta_{\sigma_{3},\downarrow}\delta_{\sigma_{4},\uparrow}W_{\downarrow\uparrow}(2,3)]\\
&\delta(\tau_{3}-\tau_{4})\delta(\tau_{1}-\tau_{2})\delta_{\boldsymbol{R}_{3},\boldsymbol{R}_{4}}\delta_{\boldsymbol{R}_{1},\boldsymbol{R}_{2}}.
\end{aligned}
\end{equation}
One can visualize the screened interactions as

\begin{center}
\begin{tikzpicture}
    \coordinate (A) at (0,0.8);    
    \coordinate (B) at (2.5,0.8);    
    \coordinate (C) at (0,0);    
    \coordinate (D) at (2.5,0);    
    
    \draw (A) -- (B);  
    \draw (C) -- (D);  
    
    \draw[decorate, decoration={zigzag, segment length=0.2cm, amplitude=0.07cm}] (A) -- (C);  
    \draw[decorate, decoration={zigzag, segment length=0.2cm, amplitude=0.07cm}] (B) -- (D);  
    
    \node[fill=white, inner sep=1pt] at (A) {$\uparrow$\;4};
    \node[fill=white, inner sep=1pt] at (B) {1\;$\uparrow$};
    \node[fill=white, inner sep=1pt] at (C) {$\downarrow$\;3};
    \node[fill=white, inner sep=1pt] at (D) {2\;$\downarrow$};
    
    \node at (3.5,0.4) {$+$};
    
    \coordinate (A2) at (4.5,0.8);    
    \coordinate (B2) at (7.0,0.8);    
    \coordinate (C2) at (4.5,0);    
    \coordinate (D2) at (7.0,0);    
    
    \draw (A2) -- (B2);  
    \draw (C2) -- (D2);  
    
    \draw[decorate, decoration={zigzag, segment length=0.2cm, amplitude=0.07cm}] (A2) -- (C2);  
    \draw[decorate, decoration={zigzag, segment length=0.2cm, amplitude=0.07cm}] (B2) -- (D2);  
    
    \node[fill=white, inner sep=1pt] at (A2) {$\downarrow$\;4};
    \node[fill=white, inner sep=1pt] at (B2) {1\;$\downarrow$};
    \node[fill=white, inner sep=1pt] at (C2) {$\uparrow$\;3};
    \node[fill=white, inner sep=1pt] at (D2) {2\;$\uparrow$};
\end{tikzpicture}
\end{center}

With the interaction vertices being $\sum_{n=3}^{\infty}\frac{i^n}{n}\mathrm{Tr}[(G_{*}\delta\Sigma)^n]$, all the Feynman diagrams correspond to the sum of perfect matching with fermion lines partitioned into different fermion loops. \textit{E.g.} at order six, there are nine partitions with minimal number being $3$: $(12,)$, $(9,3)$, $(8,4)$, $(7,5)$, $(6,6)$, $(6,3,3)$, $(5,4,3)$, $(4,4,4)$, and $(3,3,3,3)$.
For example, for $(6,6)$, two Feynman diagrams can be
\begin{center}
\begin{tikzpicture}
  \def\radius{1.0}
  \def\vshift{2.8} 
  \def\hshift{3.6} 
  
  
  \node[circle, fill=black, inner sep=0.5pt] (v0) at (0:\radius) {};
  \node[circle, fill=black, inner sep=0.5pt] (v1) at (10:\radius) {};
  \node[font=\tiny] at (0:\radius+0.3) {0};
  \node[font=\tiny] at (10:\radius+0.3) {1};
  
  \node[circle, fill=black, inner sep=0.5pt] (v2) at (60:\radius) {};
  \node[circle, fill=black, inner sep=0.5pt] (v3) at (70:\radius) {};
  \node[font=\tiny] at (60:\radius+0.3) {2};
  \node[font=\tiny] at (70:\radius+0.3) {3};
  
  \node[circle, fill=black, inner sep=0.5pt] (v4) at (120:\radius) {};
  \node[circle, fill=black, inner sep=0.5pt] (v5) at (130:\radius) {};
  \node[font=\tiny] at (120:\radius+0.3) {4};
  \node[font=\tiny] at (130:\radius+0.3) {5};
  
  \node[circle, fill=black, inner sep=0.5pt] (v6) at (180:\radius) {};
  \node[circle, fill=black, inner sep=0.5pt] (v7) at (190:\radius) {};
  \node[font=\tiny] at (180:\radius+0.3) {6};
  \node[font=\tiny] at (190:\radius+0.3) {7};
  
  \node[circle, fill=black, inner sep=0.5pt] (v8) at (240:\radius) {};
  \node[circle, fill=black, inner sep=0.5pt] (v9) at (250:\radius) {};
  \node[font=\tiny] at (230:\radius-0.2) {8};
  \node[font=\tiny] at (250:\radius-0.2) {9};
  
  \node[circle, fill=black, inner sep=0.5pt] (v10) at (280:\radius) {};
  \node[circle, fill=black, inner sep=0.5pt] (v11) at (290:\radius) {};
  \node[font=\tiny] at (275:\radius-0.2) {10};
  \node[font=\tiny] at (300:\radius-0.2) {11};
  
  
  \draw[blue, opacity=0.3, line width=2pt] (290:\radius) arc (290:360:\radius);
  
  \draw (10:\radius) arc (10:60:\radius);
  \draw (70:\radius) arc (70:120:\radius);
  \draw (130:\radius) arc (130:180:\radius);
  \draw (190:\radius) arc (190:240:\radius);
  \draw (250:\radius) arc (250:280:\radius);
  \draw (290:\radius) arc (290:360:\radius);
  
  \draw[thick] (v1) to[bend left=30] (v2);
  \draw[thick, dashed] (v2) to[bend right=30] (v1);
  
  \draw[blue, opacity=0.3, line width=2.5pt] (v3) to[bend right=30] (v0);
  \draw[thick] (v3) to[bend right=30] (v0);
  
  \draw[blue, opacity=0.3, line width=2.5pt] (v0) -- (v1);
  \draw[blue, opacity=0.3, line width=2.5pt] (v1) to[bend left=30] (v2);
   \draw[blue, opacity=0.3, line width=2.5pt] (v1) to[bend right=20] (v2);
  \draw[blue, opacity=0.3, line width=2.5pt] (v2) -- (v3);
  
  \draw[decorate, decoration={zigzag, amplitude=1.5pt, segment length=2pt}] (v0) -- (v1);
  \draw[decorate, decoration={zigzag, amplitude=1.5pt, segment length=2pt}] (v2) -- (v3);
  \draw[decorate, decoration={zigzag, amplitude=1.5pt, segment length=2pt}] (v4) -- (v5);
  \draw[decorate, decoration={zigzag, amplitude=1.5pt, segment length=2pt}] (v6) -- (v7);
  \draw[decorate, decoration={zigzag, amplitude=1.5pt, segment length=2pt}] (v8) -- (v9);
  \draw[decorate, decoration={zigzag, amplitude=1.5pt, segment length=2pt}] (v10) -- (v11);
    
  \draw[thick] (v4) to[bend left=40] (v7);
  
  \draw[thick] (v5) to[bend left=40] (v6);
    
  
  \coordinate (bottom_center) at (0,-\vshift);
  
  \node[circle, fill=black, inner sep=0.5pt] (v12) at ($(bottom_center) + (75:\radius)$) {};
  \node[circle, fill=black, inner sep=0.5pt] (v13) at ($(bottom_center) + (85:\radius)$) {};
  \node[font=\tiny] at ($(bottom_center) + (65:\radius-0.2)$) {12};
  \node[font=\tiny] at ($(bottom_center) + (85:\radius-0.2)$) {13};
  
  \node[circle, fill=black, inner sep=0.5pt] (v14) at ($(bottom_center) + (110:\radius)$) {};
  \node[circle, fill=black, inner sep=0.5pt] (v15) at ($(bottom_center) + (120:\radius)$) {};
  \node[font=\tiny] at ($(bottom_center) + (110:\radius-0.2)$) {14};
  \node[font=\tiny] at ($(bottom_center) + (130:\radius-0.2)$) {15};
  
  \node[circle, fill=black, inner sep=0.5pt] (v16) at ($(bottom_center) + (180:\radius)$) {};
  \node[circle, fill=black, inner sep=0.5pt] (v17) at ($(bottom_center) + (190:\radius)$) {};
  \node[font=\tiny] at ($(bottom_center) + (180:\radius+0.3)$) {16};
  \node[font=\tiny] at ($(bottom_center) + (190:\radius+0.3)$) {17};
  
  \node[circle, fill=black, inner sep=0.5pt] (v18) at ($(bottom_center) + (240:\radius)$) {};
  \node[circle, fill=black, inner sep=0.5pt] (v19) at ($(bottom_center) + (250:\radius)$) {};
  \node[font=\tiny] at ($(bottom_center) + (240:\radius+0.3)$) {18};
  \node[font=\tiny] at ($(bottom_center) + (250:\radius+0.3)$) {19};
  
  \node[circle, fill=black, inner sep=0.5pt] (v21) at ($(bottom_center) + (300:\radius)$) {};
  \node[circle, fill=black, inner sep=0.5pt] (v20) at ($(bottom_center) + (310:\radius)$) {};
  \node[font=\tiny] at ($(bottom_center) + (300:\radius+0.3)$) {21};
  \node[font=\tiny] at ($(bottom_center) + (310:\radius+0.3)$) {20};
  
  \node[circle, fill=black, inner sep=0.5pt] (v22) at ($(bottom_center) + (0:\radius)$) {};
  \node[circle, fill=black, inner sep=0.5pt] (v23) at ($(bottom_center) + (10:\radius)$) {};
  \node[font=\tiny] at ($(bottom_center) + (0:\radius+0.3)$) {22};
  \node[font=\tiny] at ($(bottom_center) + (10:\radius+0.3)$) {23};
  
  \draw[decorate, decoration={zigzag, amplitude=1.5pt, segment length=2pt}] (v12) -- (v13);
  \draw[decorate, decoration={zigzag, amplitude=1.5pt, segment length=2pt}] (v14) -- (v15);
  \draw[decorate, decoration={zigzag, amplitude=1.5pt, segment length=2pt}] (v16) -- (v17);
  \draw[decorate, decoration={zigzag, amplitude=1.5pt, segment length=2pt}] (v18) -- (v19);
  \draw[decorate, decoration={zigzag, amplitude=1.5pt, segment length=2pt}] (v21) -- (v20);
  \draw[decorate, decoration={zigzag, amplitude=1.5pt, segment length=2pt}] (v22) -- (v23);
  
  \draw ($(bottom_center) + (85:\radius)$) arc (85:110:\radius);
  \draw ($(bottom_center) + (120:\radius)$) arc (120:180:\radius);
  \draw ($(bottom_center) + (190:\radius)$) arc (190:240:\radius);
  \draw ($(bottom_center) + (250:\radius)$) arc (250:300:\radius);
  \draw ($(bottom_center) + (310:\radius)$) arc (310:360:\radius);
  \draw ($(bottom_center) + (10:\radius)$) arc (10:75:\radius);

  \draw[thick] (v8) to[bend left=0] (v13);
  \draw[thick] (v9) to[bend left=0] (v12);
  \draw[thick] (v10) to[bend left=0] (v15);
  \draw[thick] (v11) to[bend left=0] (v14);

  \draw[thick] (v17) to[bend left=30] (v18);
  \draw[thick] (v16) to[bend left=30] (v19);
  \draw[thick] (v23) to[bend right=30] (v21);
  \draw[thick] (v22) to[bend right=30] (v20);

  \draw[red, opacity=0.3, line width=2.5pt] (v4) --(v5);
  \draw[red, opacity=0.3, line width=2.5pt] (v6) --(v7);
  \draw[red, opacity=0.3, line width=2.5pt] (v4) to[bend left=40] (v7);
  \draw[red, opacity=0.3, line width=2.5pt] (v5) to[bend left=40] (v6);
  \draw[red, opacity=0.3, line width=2.5pt] (v5) to[bend right=20] (v6);
  \draw[red, opacity=0.3, line width=2.5pt] (v3) to[bend right=20] (v4);

  \draw[yellow, opacity=0.3, line width=2.5pt] (v8) --(v9);
  \draw[yellow, opacity=0.3, line width=2.5pt] (v13) --(v12);
  \draw[yellow, opacity=0.3, line width=2.5pt] (v7) to[bend right=20] (v8);
  \draw[yellow, opacity=0.3, line width=2.5pt] (v9) to[bend right=20] (v10);
  \draw[yellow, opacity=0.3, line width=2.5pt] (v8) --(v13);
  \draw[yellow, opacity=0.3, line width=2.5pt] (v9) --(v12);

  \draw[gray, opacity=0.3, line width=2.5pt] (v10) --(v11);
  \draw[gray, opacity=0.3, line width=2.5pt] (v15) --(v14);
  \draw[gray, opacity=0.3, line width=2.5pt] (v16) to[bend left=25] (v15);
  \draw[gray, opacity=0.3, line width=2.5pt] (v14) to[bend left=15] (v13);
  \draw[gray, opacity=0.3, line width=2.5pt] (v10) --(v15);
  \draw[gray, opacity=0.3, line width=2.5pt] (v11) --(v14);

  \definecolor{mygreen}{rgb}{0.1, 0.9, 0.1}
  \draw[mygreen, opacity=0.3, line width=2.5pt] (v16) --(v17);
  \draw[mygreen, opacity=0.3, line width=2.5pt] (v18) --(v19);  
  \draw[mygreen, opacity=0.3, line width=2.5pt] (v17) to[bend right=20](v18); 
  \draw[mygreen, opacity=0.3, line width=2.5pt] (v19) to[bend right=20](v21);  
  \draw[mygreen, opacity=0.3, line width=2.5pt] (v17) to[bend left=30](v18);
  \draw[mygreen, opacity=0.3, line width=2.5pt] (v16) to[bend left=30](v19); 

  \definecolor{mybrown}{rgb}{0.6, 0.17, 0.17}
  \draw[mybrown, opacity=0.3, line width=2.5pt] (v21) --(v20);
  \draw[mybrown, opacity=0.3, line width=2.5pt] (v22) --(v23);  
  \draw[mybrown, opacity=0.3, line width=2.5pt] (v20) to[bend right=20](v22); 
  \draw[mybrown, opacity=0.3, line width=2.5pt] (v23) to[bend right=27](v12);  
  \draw[mybrown, opacity=0.3, line width=2.5pt] (v20) to[bend left=30](v22);
  \draw[mybrown, opacity=0.3, line width=2.5pt] (v21) to[bend left=30](v23); 

  
  \coordinate (right_top_center) at (\hshift,0);
  
  \node[circle, fill=black, inner sep=0.5pt] (v24) at ($(right_top_center) + (0:\radius)$) {};
  \node[circle, fill=black, inner sep=0.5pt] (v25) at ($(right_top_center) + (10:\radius)$) {};
  \node[font=\tiny] at ($(right_top_center) + (0:\radius+0.3)$) {0};
  \node[font=\tiny] at ($(right_top_center) + (10:\radius+0.3)$) {1};
  
  \node[circle, fill=black, inner sep=0.5pt] (v26) at ($(right_top_center) + (60:\radius)$) {};
  \node[circle, fill=black, inner sep=0.5pt] (v27) at ($(right_top_center) + (70:\radius)$) {};
  \node[font=\tiny] at ($(right_top_center) + (60:\radius+0.3)$) {2};
  \node[font=\tiny] at ($(right_top_center) + (70:\radius+0.3)$) {3};
  
  \node[circle, fill=black, inner sep=0.5pt] (v28) at ($(right_top_center) + (120:\radius)$) {};
  \node[circle, fill=black, inner sep=0.5pt] (v29) at ($(right_top_center) + (130:\radius)$) {};
  \node[font=\tiny] at ($(right_top_center) + (120:\radius+0.3)$) {4};
  \node[font=\tiny] at ($(right_top_center) + (130:\radius+0.3)$) {5};
  
  \node[circle, fill=black, inner sep=0.5pt] (v30) at ($(right_top_center) + (180:\radius)$) {};
  \node[circle, fill=black, inner sep=0.5pt] (v31) at ($(right_top_center) + (190:\radius)$) {};
  \node[font=\tiny] at ($(right_top_center) + (180:\radius+0.3)$) {6};
  \node[font=\tiny] at ($(right_top_center) + (190:\radius+0.3)$) {7};
  
  \node[circle, fill=black, inner sep=0.5pt] (v32) at ($(right_top_center) + (240:\radius)$) {};
  \node[circle, fill=black, inner sep=0.5pt] (v33) at ($(right_top_center) + (250:\radius)$) {};
  \node[font=\tiny] at ($(right_top_center) + (230:\radius-0.2)$) {8};
  \node[font=\tiny] at ($(right_top_center) + (250:\radius-0.2)$) {9};
  
  \node[circle, fill=black, inner sep=0.5pt] (v34) at ($(right_top_center) + (280:\radius)$) {};
  \node[circle, fill=black, inner sep=0.5pt] (v35) at ($(right_top_center) + (290:\radius)$) {};
  \node[font=\tiny] at ($(right_top_center) + (275:\radius-0.2)$) {10};
  \node[font=\tiny] at ($(right_top_center) + (300:\radius-0.2)$) {11};

  \draw ($(right_top_center) + (10:\radius)$) arc (10:60:\radius);
  \draw ($(right_top_center) + (70:\radius)$) arc (70:120:\radius);
  \draw ($(right_top_center) + (130:\radius)$) arc (130:180:\radius);
  \draw ($(right_top_center) + (190:\radius)$) arc (190:240:\radius);
  \draw ($(right_top_center) + (250:\radius)$) arc (250:280:\radius);
  \draw ($(right_top_center) + (290:\radius)$) arc (290:360:\radius);
  
  \draw[thick] (v25) to[bend left=30] (v28);
  \draw[thick] (v24) to[bend left=30] (v29);
  
  \draw[decorate, decoration={zigzag, amplitude=1.5pt, segment length=2pt}] (v24) -- (v25);
  \draw[decorate, decoration={zigzag, amplitude=1.5pt, segment length=2pt}] (v26) -- (v27);
  \draw[decorate, decoration={zigzag, amplitude=1.5pt, segment length=2pt}] (v28) -- (v29);
  \draw[decorate, decoration={zigzag, amplitude=1.5pt, segment length=2pt}] (v30) -- (v31);
  \draw[decorate, decoration={zigzag, amplitude=1.5pt, segment length=2pt}] (v32) -- (v33);
  \draw[decorate, decoration={zigzag, amplitude=1.5pt, segment length=2pt}] (v34) -- (v35);
    
  \draw[thick] (v30) to[bend right=40] (v27);
  \draw[thick] (v31) to[bend right=40] (v26);
    
  
  \coordinate (right_bottom_center) at (\hshift,-\vshift);
  
  \node[circle, fill=black, inner sep=0.5pt] (v36) at ($(right_bottom_center) + (75:\radius)$) {};
  \node[circle, fill=black, inner sep=0.5pt] (v37) at ($(right_bottom_center) + (85:\radius)$) {};
  \node[font=\tiny] at ($(right_bottom_center) + (65:\radius-0.2)$) {12};
  \node[font=\tiny] at ($(right_bottom_center) + (85:\radius-0.2)$) {13};
  
  \node[circle, fill=black, inner sep=0.5pt] (v38) at ($(right_bottom_center) + (110:\radius)$) {};
  \node[circle, fill=black, inner sep=0.5pt] (v39) at ($(right_bottom_center) + (120:\radius)$) {};
  \node[font=\tiny] at ($(right_bottom_center) + (110:\radius-0.2)$) {14};
  \node[font=\tiny] at ($(right_bottom_center) + (130:\radius-0.2)$) {15};
  
  \node[circle, fill=black, inner sep=0.5pt] (v40) at ($(right_bottom_center) + (180:\radius)$) {};
  \node[circle, fill=black, inner sep=0.5pt] (v41) at ($(right_bottom_center) + (190:\radius)$) {};
  \node[font=\tiny] at ($(right_bottom_center) + (180:\radius+0.3)$) {16};
  \node[font=\tiny] at ($(right_bottom_center) + (190:\radius+0.3)$) {17};
  
  \node[circle, fill=black, inner sep=0.5pt] (v42) at ($(right_bottom_center) + (240:\radius)$) {};
  \node[circle, fill=black, inner sep=0.5pt] (v43) at ($(right_bottom_center) + (250:\radius)$) {};
  \node[font=\tiny] at ($(right_bottom_center) + (240:\radius+0.3)$) {18};
  \node[font=\tiny] at ($(right_bottom_center) + (250:\radius+0.3)$) {19};
  
  \node[circle, fill=black, inner sep=0.5pt] (v45) at ($(right_bottom_center) + (300:\radius)$) {};
  \node[circle, fill=black, inner sep=0.5pt] (v44) at ($(right_bottom_center) + (310:\radius)$) {};
  \node[font=\tiny] at ($(right_bottom_center) + (300:\radius+0.3)$) {21};
  \node[font=\tiny] at ($(right_bottom_center) + (310:\radius+0.3)$) {20};
  
  \node[circle, fill=black, inner sep=0.5pt] (v46) at ($(right_bottom_center) + (0:\radius)$) {};
  \node[circle, fill=black, inner sep=0.5pt] (v47) at ($(right_bottom_center) + (10:\radius)$) {};
  \node[font=\tiny] at ($(right_bottom_center) + (0:\radius+0.3)$) {22};
  \node[font=\tiny] at ($(right_bottom_center) + (10:\radius+0.3)$) {23};
  
  \draw[decorate, decoration={zigzag, amplitude=1.5pt, segment length=2pt}] (v36) -- (v37);
  \draw[decorate, decoration={zigzag, amplitude=1.5pt, segment length=2pt}] (v38) -- (v39);
  \draw[decorate, decoration={zigzag, amplitude=1.5pt, segment length=2pt}] (v40) -- (v41);
  \draw[decorate, decoration={zigzag, amplitude=1.5pt, segment length=2pt}] (v42) -- (v43);
  \draw[decorate, decoration={zigzag, amplitude=1.5pt, segment length=2pt}] (v45) -- (v44);
  \draw[decorate, decoration={zigzag, amplitude=1.5pt, segment length=2pt}] (v46) -- (v47);
  
  \draw ($(right_bottom_center) + (85:\radius)$) arc (85:110:\radius);
  \draw ($(right_bottom_center) + (120:\radius)$) arc (120:180:\radius);
  \draw ($(right_bottom_center) + (190:\radius)$) arc (190:240:\radius);
  \draw ($(right_bottom_center) + (250:\radius)$) arc (250:300:\radius);
  \draw ($(right_bottom_center) + (310:\radius)$) arc (310:360:\radius);
  \draw ($(right_bottom_center) + (10:\radius)$) arc (10:75:\radius);

  \draw[thick] (v32) to[bend left=0] (v37);
  \draw[thick] (v33) to[bend left=0] (v36);
  \draw[thick] (v34) to[bend left=0] (v39);
  \draw[thick] (v35) to[bend left=0] (v38);

  \draw[thick] (v41) to[bend left=30] (v42);
  \draw[thick] (v40) to[bend left=30] (v43);
  \draw[thick] (v47) to[bend right=30] (v45);
  \draw[thick] (v46) to[bend right=30] (v44);

  \draw[blue, opacity=0.3, line width=2.5pt] (v24) --(v25);
  \draw[blue, opacity=0.3, line width=2.5pt] (v26) --(v27);
  \draw[blue, opacity=0.3, line width=2.5pt] (v25) to[bend right=20] (v26);
  \draw[blue, opacity=0.3, line width=2.5pt] (v27) to[bend right=20] (v28);
  \draw[blue, opacity=0.3, line width=2.5pt] (v27) to[bend left=40] (v30);  
  \draw[blue, opacity=0.3, line width=2.5pt] (v26) to[bend left=40] (v31);  
  
  \draw[yellow, opacity=0.3, line width=2.5pt] (v32) --(v33);
  \draw[yellow, opacity=0.3, line width=2.5pt] (v37) --(v36);
  \draw[yellow, opacity=0.3, line width=2.5pt] (v31) to[bend right=20] (v32);
  \draw[yellow, opacity=0.3, line width=2.5pt] (v33) to[bend right=20] (v34);
  \draw[yellow, opacity=0.3, line width=2.5pt] (v32) --(v37);
  \draw[yellow, opacity=0.3, line width=2.5pt] (v33) --(v36);

  \draw[gray, opacity=0.3, line width=2.5pt] (v34) --(v35);
  \draw[gray, opacity=0.3, line width=2.5pt] (v39) --(v38);
  \draw[gray, opacity=0.3, line width=2.5pt] (v40) to[bend left=25] (v39);
  \draw[gray, opacity=0.3, line width=2.5pt] (v38) to[bend left=15] (v37);
  \draw[gray, opacity=0.3, line width=2.5pt] (v34) --(v39);
  \draw[gray, opacity=0.3, line width=2.5pt] (v35) --(v38);

  \draw[mygreen, opacity=0.3, line width=2.5pt] (v40) --(v41);
  \draw[mygreen, opacity=0.3, line width=2.5pt] (v42) --(v43);  
  \draw[mygreen, opacity=0.3, line width=2.5pt] (v41) to[bend right=20](v42); 
  \draw[mygreen, opacity=0.3, line width=2.5pt] (v43) to[bend right=20](v45);  
  \draw[mygreen, opacity=0.3, line width=2.5pt] (v41) to[bend left=30](v42);
  \draw[mygreen, opacity=0.3, line width=2.5pt] (v40) to[bend left=30](v43); 

  \draw[mybrown, opacity=0.3, line width=2.5pt] (v45) --(v44);
  \draw[mybrown, opacity=0.3, line width=2.5pt] (v46) --(v47);  
  \draw[mybrown, opacity=0.3, line width=2.5pt] (v44) to[bend right=20](v46); 
  \draw[mybrown, opacity=0.3, line width=2.5pt] (v47) to[bend right=27](v36);  
  \draw[mybrown, opacity=0.3, line width=2.5pt] (v44) to[bend left=30](v46);
  \draw[mybrown, opacity=0.3, line width=2.5pt] (v45) to[bend left=30](v47); 
  
\end{tikzpicture}
\end{center}
We have covered one screened interaction and two Green's function with colored $T-$shape cover, where 
\begin{equation}
T(1,3;3')=\sumint_{2} G_{*}(1,2)G_{*}(2,3)W(2,3'). \label{eq:T_shape_function}
\end{equation}
If the bare interaction is purely contact, this function can avoid contact/non-contact interaction switches. Given the discrete Lehmann representation for $W$ and $G_{*}$, the computation cost for each such function in real space \cite{shi2022one_over_N} only grows quadratically in the number of basis exponential functions, and grow linearly in momentum space \cite{shi2022one_over_N}. From now on, we collapse vertices in $W$ with delta functions and represent them as bold red wiggle lines. 

The left diagram admits an eligible spin configurations assuming that off-diagonal spin elements in $G_{*,\sigma\sigma'}(\tau-\tau',\boldsymbol{R}_i,\boldsymbol{R}_j)$ are zero. This is usually the case when solving the self-consistent saddle-point equations. The right diagram does not, and there cannot be a full $T-$shape cover. We verified that for all the diagrams that admit a spin-allowed configuration, there always exist a full $T-$shape cover. For a given diagram, there is an efficient algorithm to cover it with as many $T-$shapes as it could admit. We postpone the presentation of the algorithm after introducing the dynamical programming idea for Hafnian. It is helpful to investigate properties of spin alignments for a given diagram, where the following theorem is helpful.
\begin{theorem}
If a diagram admits one allowed spin configuration, then it admits two and only two allowed spin configurations.
\end{theorem}

The proof is simple just from the definition of connectedness. Since if one spin configuration is allowed, filling all the spins would generate another. For the only two part, consider a diagram with $m$ closed fermion loops. Since given spins for one vertex within a loop would determine all the others within a loop, we are left with $2^m$ possibilities. We denote the allowed spin configuration as $x_{1}...x_{m}$, where $x_{i}\in\{A,B\}$ denotes spin orientation for the given loop $i$. If one flips the orientation for any $0<p<m$ loops, then there is always a loop not being flipped such that a screened interaction emanating from it that admits the disallowed spin configuration. Therefore, either one does not flip any or one flips all. Hence only two are allowed.  

We use the following to tell if a given diagram admits spin-allowed configurations or not. Starting from an arbitrary loop and an arbitrary vertex with an arbitrary spin assignment, one performs the depth-first search (DFS) to determine the order of visits of the preceding loop. Spins are then assigned for all the vertices; and hence determining spin-allowed or not.

\textit{Combinatorial Structures with Contact Hubbard Interactions on Bravis Lattices}-- The combinatorial structure of summing all the Feynman diagrams resembles that of the Hafnians. For a given set of $2n$ vertices, we connect them with edges with adjacency matrix $A$ such that each vertex is connected with one and only one edge (\text{i.e.} sum of all perfect matching in a complete graph). The edge values are drawn from the adjacency matrix $A$ and the final expression read (normalization factor omitted compared with standard Hafnian definition)
\begin{equation}
\operatorname{haf}(A)=\sum_{\sigma \in S_{2 n}} \prod_{i=1}^n A_{\sigma(2 i-1), \sigma(2 i)},
\end{equation}
where $S_{2n}$ is the symmetry group of ${1,...,2n}$. And there is a well-known Ryser's formula that can compute it in $\mathcal{O}(n^32^n)$ \cite{hafnian_computation}.

Our case is similar, though not identical to the Hafnian computations. Firstly, there are many diagrams are filtered out due to spin-restrictions. Secondly, for each screened interaction in the diagram, it may have multiple different $T-$shape territories, that the adjacency matrix is expanded by tenor product $A_{ij}\otimes\mathcal{R}$ to record this additional information. We find that the new $\mathcal{R}$ does not spawn the nodes heavily, where there are at most two different $T-$shapes that share the same screened interaction after covering. 

Before presenting the algorithm, we digest the original dynamical programming if the structure forms a perfect Halfnian. For the summation of all perfect matchings at order $n$, we employ bitmask representation denote each node with $2n$ numbers with $0$ and $1$. Each node represent summation of some subgraphs with $1$ denoting positions that vertices are already connected by an edge. Nodes are organized into layer. The first layer only has one node which is $\boxed{00...0}$. The last layer only has one node which is $\boxed{11...1}$. There are $n$ layers in total. The first $j$ digits for all nodes in the $j^{\mathrm{th}}$ ($j=0,...,n-1$) layer are all 1's. Nodes at the $j^{\mathrm{th}}$ layer have $2j$ 1's in total. Inter-layer connection are defined as: a node B in the $(j+1)^{\mathrm{th}}$ layer is connected to a node A in the $j^{\mathrm{th}}$ layer if $0$'s in B are in $0$'s in A; $1$'s in A are in $1$'s in B; The first position for $1$'s in B that are not in A must start from the first position of $0$ in A. Number of edges (=number of multiplications) has good scalability between $\mathcal{O}(n^32^n)$ and $\mathcal{O}(3^n)$, where there are $6$, $26$, $97$, $332$, $1076$, $3361$, $10226$, $30510$, $89665$, $260376$, $748776$, $2136001$, $6052062$, $17048642$ of them from orders two to fifteen.

One can observe that the key of the construction is the rule: first position for $1$'s in B that are not in A must start from the first position of $0$ in A. In plain words, interactions in all the sub-graphs are ordered with such no-skipping rule. For $n=3$ as the example, let the vertices be $V=\{0,1,2,3,4,5\}$. The first interaction line must connect $0,i$ for $i\geq1$. For the second interaction, $j, k$. $j$ must start with the first position that is not connected yet, \text{e.g.} if $i=2$, then $j=1$. Afterwards, $0,i,j,k$ are removed from the vertices set and the third interaction must connect the minimal in $V\setminus\{0,i,j,k\}$ to anything else. This procedure assigns an ordering of all the interactions such that sub-nodes are distributed ``equally" and with high re-usability. This is the key to ensure good scalability compared to common sub-expression eliminations or Hörner's schemes since the latter would generate large nodes that prevent from sufficient factorization. The layered directed acyclic graph (later on referred to as call graph for abbreviation) building nodes connections at $n=3$ is shown in Fig. 1 (a).  
\onecolumngrid
\begin{figure}[H]
\begin{center}
\includegraphics[width=0.9\textwidth]{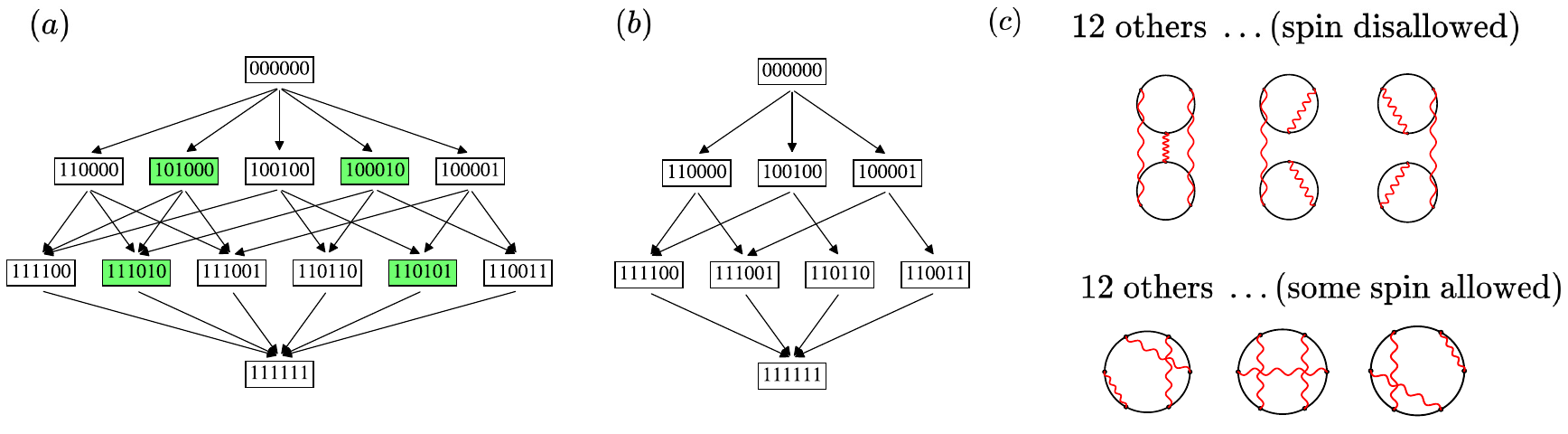}
\twocolcaption{FIG. 1. Bold-line Feynman diagrams with three screened interactions and the corresponding call graph. We present modifications of the call graphs following the main text, where green nodes are removed due to spin restrictions. At order three, there are no nodes with extra labels but this is not general. We give a full order four call graph in the End Matter, where we find very useful to demonstrate the algorithm. There are two different partitions $(6,)$ and $(3, 3)$, where $(3,3)$ is completely spin disallowed. $0$'s in the call graph nodes represent vertices that are not connected by a screened interaction while $1's$ represent vertices that are already connected by a screened interaction. Each edge is a $T-$shape function.}
\label{Fig:dp_call_graph}
\end{center}
\end{figure}
\twocolumngrid
Ideas of dynamical programming for sum of perfect matchings are classical textbook results which can be found in common graph algorithm texts. Before presenting the algorithm, we first try to cover the diagram with as many $T-$shape functions as well. We find an efficient algorithm, outlined in Alg. \ref{alg:max_T_cover}. 
\begin{figure}[ht]
    \centering
  \begin{minipage}{0.95\linewidth}

    \rule{\linewidth}{0.4pt}
    \vspace{-1.2em}

    \centering
    \textbf{Algorithm 1:} Find the maximal $T$-cover

    \vspace{-0.6em}

    \rule{\linewidth}{0.4pt}

    \begin{algorithmic}[1]
      \State $\text{graph} \gets W's\;\text{with\;no\;skipping rule}$
      \State $N \gets \#W's$
      \State $n \gets 0$
      \State $\text{partition} \gets (p_0,p_1,p_2,...)$
      \State $T_c \gets []$
      \While{$n \leq N-1$}
      \State $W \gets \text{current W}$
      \If{$n=0$}
          \State $T_c\gets T_c \cup [[p_0-1,0],[0,1],W]$
      \ElsIf{$W[0],\;W[1]\;\text{are free}$}
      
      \If{$\exists G_{*}'s,\;s.t.\; G_{*,0}\cap\mathcal{C}\neq\varnothing\lor G_{*,1}\cap\mathcal{C}\neq\varnothing$}
              \State $T_c\gets T_c \cup [G_{*,0},G_{*,1},W]$ 
              \Else
              \State $T_c\gets T_c \cup [\forall(G_{*,0}, G_{*,1}),W]$ 
          \EndIf
      \ElsIf{$W[0]\text{\;is\;free}\lor W[1]\text{\;is\;free}$}
      
      \State $[G_{*,0},G_{*,1}]\gets G_{*}'s\text{\;of\;free\;end}$
      \State $T_c\gets T_c \cup[G_{*,0},G_{*,1},W]$ 
      \Else
      \State \textbf{continue}
      \EndIf
      \State $n \leftarrow n+1$
      \EndWhile
    \end{algorithmic}

    \rule{\linewidth}{0.4pt}
  \end{minipage}
  \setcounter{figure}{0}    \refstepcounter{figure}
  \label{alg:max_T_cover}
\end{figure}
$\mathcal{C}$ is the current territory made of occupied Green's functions, which is defined as
\begin{definition}\label{def:territory}
A territory is a collection of non-repeated Green's functions edges.
\end{definition}
$T_c$ is the current cover of $T-$shape functions. For any $W$, it has two ends, where at each end there are two $G_{*}'s$. The free end of $W$ denotes the end that has two $G_{*}'s$ are are still not covered. There is some freedom when both ends of the current $W$ are not occupied and none of the two $G_{*}'s$ at neither ends of $W$ is contact with the current territory. This case happens when the current screened interaction starts a new cycle. And we simply try for multiple times and we found for at most two attempts the cover becomes full. The key for fast success is to start with ordered $W$'s in the sense of no-skipping rule. This has dual impact, not only for a fast $T-$cover finding, but also remains the combinatorial structure of the original Hafnian as much as possible, which then lays the foundation for the latter dynamical programming algorithm.

Additional complexities are several-fold. One has to remove the disconnected diagrams and may wish to switch to momentum space for more compact Monte-Carlo configurations. In other channels, the cover by $T-$shape functions are not full and thus one has to make contact/non-contact contractions for the remaining interactions. All of them can be solved in a unified way by just modifying the call graph with the no-skipping rule ordered interaction. 

Starting from all the graphs covered by the $T-$cover, we first group them with their last element, which form nodes in the layer above the last layer. For each node in the last layer, we group them with the last element and build nodes in the layer above. If the parent node from the child already exists in layer above the current, we simply connect the parent node with the child node; otherwise, we create that new parent node. Each layer is assigned with a hash table with the nodes contents as the key such that the look-up time to check if a parent node already exists or not is $\mathcal{O}(1)$. 

Although the complexity of the algorithm to build the call graph is factorial, this can be done fast for realistic expansion orders up to $\sim 10$. For all the parameters under sweeping (\textit{e.g.} temperature, interaction, filling), the call graph is shared in common. Compared with the Monte-Carlo sampling afterwards, required computational time is marginal even with just one parameter. 

\textit{Generalized to Hubbard Contact Interactions in Arbitrarily Lattice}--
The case with Bravis lattices are less general and in this section we generalize the construction to the most general lattices. Since given a node, we build its parent nodes by grouping the last common $T-$shape building block and construct one for each such group. Therefore, by construction, all the subgraphs within a node have the same territory (a straightforward proof by induction). For each $T-$shape function defined in Eq. (\ref{eq:T_shape_function})
we call vertices $1$, $3$, $3'$ external vertices of $T$, denoted by $\text{Ext}_{T}$ and $2$ the internal vertex. Note that $3'$ can be equal to either $1$ or $3$. For a given node, we define its external points as: 
\begin{definition}\label{def:external_pts_territory}
For a given subgraph, $F$, that are consist of $T-$shape territories, internal points are vertices that are connected by two Green's functions and one screened interaction. External points are the remaining, denoted by $\text{Ext}_{F}$. Number of external points for a given subgraph (or a given set of subgraphs if they share the same set of external points) is denoted by $\text{deg}_{F}$.
\end{definition}
Now, given that all the subgraphs within a node share the same territory, another proof by induction shows that they must also share the same set of vertices: if an external vertex in the $T-$shape territory being removed from the child node belongs to the shared territory, then that vertex is still in the vertices set; otherwise, that vertex is removed from the vertices set. From this, one can show that all the subgraphs within a node also share the same set of external points. Since from the bottommost node $\boxed{1...1}$ to nodes with two less $1's$, all the subgraphs within a node share the same set of external points. Assume that from layer $n$ to layer $(n-1)$, this holds. Let us now consider building a parent node from layer $(n-1)$ to layer $(n-2)$. Since that parent node is formed by removing one $T-$shape territory from the child node. For a given subgraph in the parent node, all the external points that are still in it must be external. And due to the removal, all the external vertices of the $T-$shape territory that are still in the parent node must be external. Therefore, all the subgraphs in the parent node must share the same set of external points. The remaining proof is thus done by induction. We summarize the above discussions as the theorem below:
\begin{theorem}\label{thm:shared_propertites}
All the subgraphs within a node share the same set of vertices, the same territory and the same set of external points.
\end{theorem}
With the definition of external points, the summation of all sub-lattice (or in the non-equilibrium set-up, the forward, backward and thermal branches of the complex time contour) can be down easily by spawning the nodes. Given that at each vertex, there is an additional $N_{s}$ degrees of freedom. For each node, we prepare the $N_{s}^{\text{deg}_{n}}$ copies of it with each external point having $N_{s}$ choices. For each incoming edge, $e$, connected to it, it proliferates to $N_{s}^{\text{deg}_{e}}$ of it. It is easy to see that $\text{deg}_{e}=|\{\text{pt}:\text{pt}\in \text{Ext}_{T},\text{pt}\notin\text{Ext}_{F}\}|$, where $T$ is that $T-$shape territory being removed from the present node and $F$ is the collection of all subgraphs in the present node. This process is visualized below, where originally we have the node 
\begin{center}
\begin{tikzpicture}
    \node[draw, rectangle, line width=1.0pt, minimum width=1.0cm, minimum height=0.3cm] (main) {node};
    
    \coordinate (target) at (main.north);
    
    \draw[->, thick, black] ($(target) + (-1.5,1.5)$) -- (target) node[pos=0.5, left] {$e_1$};
    \draw[->, thick, blue] ($(target) + (1.5,1.5)$) -- (target) node[pos=0.5, right] {$e_2$};
\end{tikzpicture}
\end{center}
This node and all its incoming edges then proliferate to
\begin{center}
\begin{tikzpicture}[scale=0.8]
    \node[draw, rectangle, line width=1pt, minimum width=1cm, minimum height=0.3cm] (node1) {node};
    \node[draw, rectangle, line width=1pt, minimum width=1cm, minimum height=0.3cm, right=3cm of node1] (node2) {node};
    
    \coordinate (target1) at (node1.north);
    \coordinate (target2) at (node2.north);
    
    \node at ($(node1)!0.5!(node2)$) {\ldots \ldots};
    
    \draw[->, thick, black] ($(target1) + (-1.6,2)$) -- (target1);
    \draw[->, thick, black] ($(target1) + (-0.6,2)$) -- (target1);
    \draw[->, thick, blue] ($(target1) + (0.8,2)$) -- (target1);
    \draw[->, thick, blue] ($(target1) + (2.0,2)$) -- (target1);
    
    \draw[->, thick, black] ($(target2) + (-1.6,2)$) -- (target2);
    \draw[->, thick, black] ($(target2) + (-0.6,2)$) -- (target2);
    \draw[->, thick, blue] ($(target2) + (0.8,2)$) -- (target2);
    \draw[->, thick, blue] ($(target2) + (2.0,2)$) -- (target2);
    
    \node at ($(target1) + (-0.9,1.8)$) {\dots};
    \node at ($(target1) + (1.25,1.8)$) {\dots};
    
    \node at ($(target2) + (-0.9,1.8)$) {\dots};
    \node at ($(target2) + (1.25,1.8)$) {\dots};
    
    \node at ($(target1) + (-1.1,2.5)$) {$N_{s}^{\text{deg}_{e_1}}$};
    \node at ($(target1) + (1.6,2.5)$) {$N_{s}^{\text{deg}_{e_2}}$};
    
    \node at ($(target2) + (-1.1,2.5)$) {$N_{s}^{\text{deg}_{e_1}}$};
    \node at ($(target2) + (1.6,2.5)$) {$N_{s}^{\text{deg}_{e_2}}$};
    
    \draw[decoration={brace,mirror,raise=5pt},decorate]
      (node1.south west) -- (node2.south east) 
      node[pos=0.5,below=10pt] {$N_s^{\deg_n}$};
\end{tikzpicture}
\end{center}
The number of edges in each call graph from orders seven to nine with number of different sub-lattices ranging from two to four is summarized in the Fig. 2. Horizontal axis denotes different partitions and the same color means they belong to the same order. Inset plot gives number of edges at orders five and six. For $N_{s}>1$, we scale the curve with $1.5^{n}$, $2.1^n$ and $2.7^n$ for visual convenience, with $n$ being the order number. Number of edges roughly follow the same curve with another set of scaling factors $1.6^n$, $2.2^n$ and $2.8^n$. This indicates that the scaling is more favorable than the state-of-art exponential scaling $N_{s}^{n}=2^n,3^n,4^n$.
\onecolumngrid
\begin{figure}[H]
\begin{center}
\includegraphics[width=1.0\textwidth]{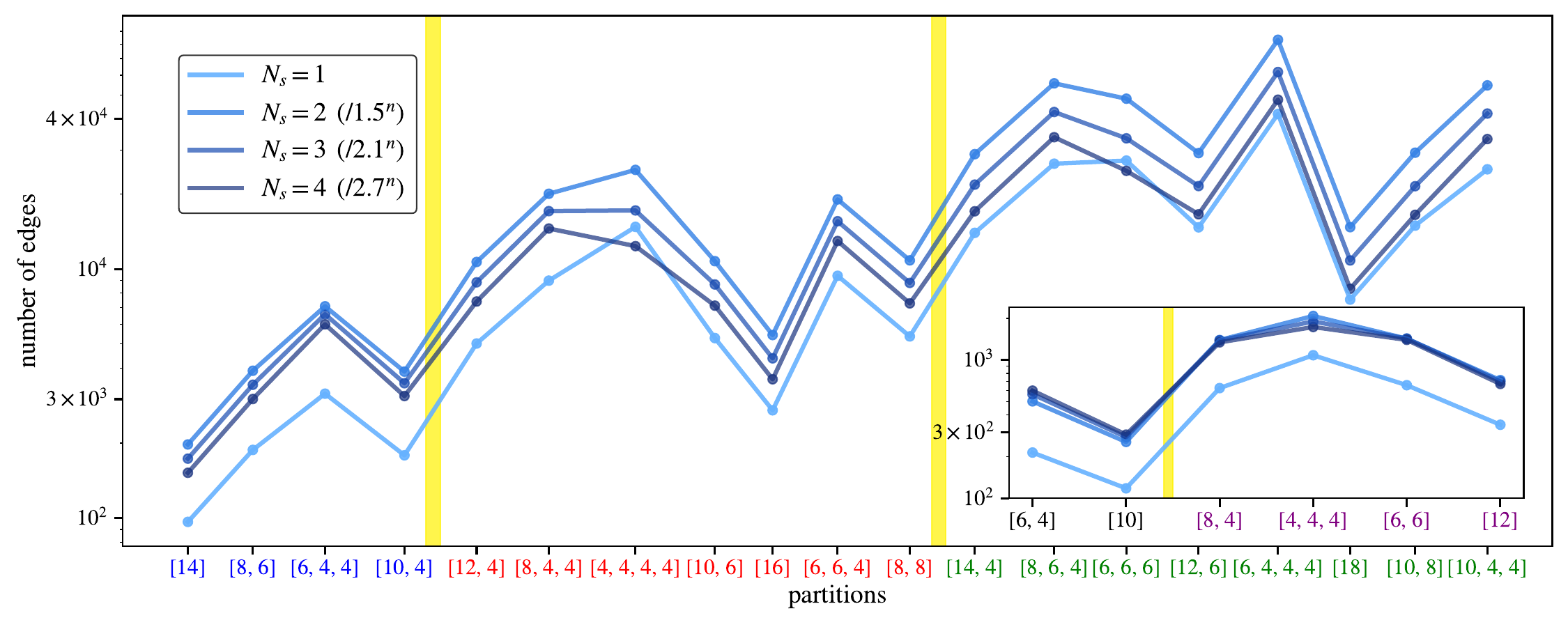}
\twocolcaption{FIG. 2: Number of edges with partitions for different number of sub-lattices. $N_{s}=1,2,3,4$ covers common lattices like the square, cubic, triangular, honeycomb, kagome, decorated square lattices, \textit{etc}. We add vertical bands to separate partitions in different orders. For a given partition, $(n_1,...,n_p)$, it has degeneracy $\prod_{i}n_{i}!/\prod_{l}c_l!$, where $c_{l}$ counts the number of occurrence for distinct numbers in a partition. For partitions up to a permutation, additional edges can be added by merging similar call graphs. The number of edges in the plot only gives one in each such family.}
\label{Fig:partitions_edges}
\end{center}
\end{figure}
\twocolumngrid
\textit{Extensions}-- There are many straightforward renormalization schemes on top of bosonized actions. One can identify common patterns in diagrams and remove the such patterns with the re-normalized Green's function. Similarly, one can further re-normalize the screened interactions. Since there are ``infinite" many of ways to do re-normalizations and terms reordering, we do not list them all here. The basic idea is that given 
\begin{equation}
\mathrm{lnZ}=\sum\text{graphs}=\sum_{n}\xi^n\;\text{group}_n|_{\xi=1},
\end{equation}
where how to group diagrams are completely arbitrary (Due to previous experiences, different groupings, subject to sufficient orders can be reached, give quite consistent results.). In particular, the $1/\mathrm{N}_f$ series in \cite{shi2022one_over_N} amounts to do power counting with $n-n_F$, where $n$ is the order number and $n_{F}$ is the number of closed fermion loop. This can be simply modified to, \textit{e.g.} $\lfloor1.5n-n_{F}\rfloor$ to make the growth of order number with this count slower.

Akin to homotopic actions, one can also consider the following action
\begin{equation}
\begin{aligned}
S(\xi)=&\sum_{n=3}^{\infty}\frac{i^n}{n}\mathrm{Tr}[(G_{*}\sqrt{\xi}\delta\Sigma)^n]+(1-\xi)\mathrm{Tr}[(G_{*}^{\prime}\delta\Sigma)^2]\\
&+\frac{1}{2}(\delta\Sigma\;\delta G)\mathcal{K}\begin{pmatrix}\delta\Sigma\\
\delta G\end{pmatrix},
\end{aligned}
\end{equation}
where $\mathcal{K}$ is the $2\times2$ bosonic propagator mentioned earlier. Expanding in $\xi$ and setting $\xi=1$ afterwards would be just adding cycles of length $2$ with modified $\mathcal{K}$ by absorbing $\mathrm{Tr}[(G_{*}'\delta\Sigma)^2]$ term. Arbitrary symmetry broken patterns can be included in $G'_{*}$ if such features are absent in the original $G_{*}$.

More advantages can be gained after forming a call graph if one employs the idea in \cite{shi2022one_over_N}, that one can use stimulated annealing to minimize the normalization integral to increase the sign by grouping diagrams. The loss function to be minimized is
\begin{equation}
\text{Loss}(Q)=C(Q)\mathcal{N}(Q)^2\tau_{f}(Q),
\end{equation}
where $Q$ is a grouping (which part of the sum is done deterministically and which part is done stochastically), $C$ is the cost for deterministic sum, $\mathcal{N}$ is the normalization integral and $\tau_{f}$ is the integrated auto-correlation time. The deterministic sum from the last to final layer, and across different partitions can be leveraged to semi-deterministic following the procedure in \cite{shi2022one_over_N}.

\textit{Conclusions and Outlook}-- We have presented a novel combinatorial platform for diagrammatic Monte Carlo that is fully versatile towards real materials simulations. Other channels and more interaction matrix elements are of course straightforward generalizations of the dynamical programming idea in this Letter, where the key is the no-skipping rule introduced in the main text.

For bold-line RPA series in the magnetic channel, we found more favorable scaling and computational costs compared with the state-of-art RPADet \cite{RPA_Det}, while our platform remains full versatility.

\textit{Acknowledgments}-- This work is supported by the Imperial College President's Scholarship. The imaginary unit $i$ of the Lagrange multipliers in the action is usually omitted in the $G-\Sigma$ literature for SYK models, and the final computational results would not change except for a sign flip in the next-to-leading order correction for $\mathrm{ln}Z$. We restore it back for more rigor.
\providecommand{\noopsort}[1]{}\providecommand{\singleletter}[1]{#1}%

\twocolumngrid  
\onecolumngrid  
\newpage
\thispagestyle{empty}
\begin{center}
    \textbf{\Large End Matter}
\end{center}
We give the modified call graph with partition $(4,4)$ in the End Matter to more intuitively demonstrate the algorithm. 
\begin{figure}[H]
\begin{center}
\includegraphics[width=0.9\textwidth]{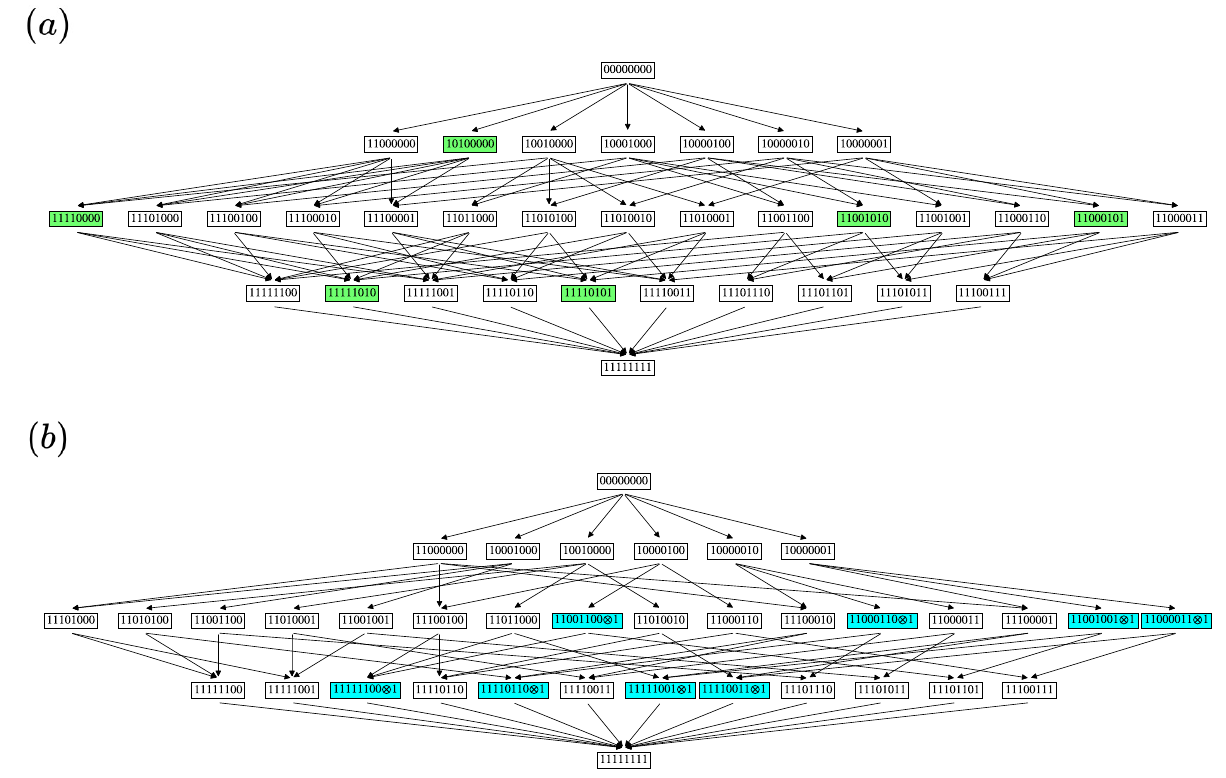}
\caption*{FIG. 3. Original and modified call graph at order $4$ for $(4, 4)$. (a): the original call graph, where green nodes are removed. (b): the modified call graph, where cyan nodes are added on top of the the original call graph.}
\label{Fig:call_graph_n_4}
\end{center}
\end{figure}
It is fairly interesting to see the correspondence between diagrams after bosonization and before. If the saddle point is trivial, \text{i.e.} $\Sigma_{*}=0$, the zeroth order in $\mathrm{ln}Z$ just corresponds to the free system. The part contributed by $\frac{1}{2}\mathrm{Tr}\mathrm{ln}[1-W_{G}W_{\Sigma}]$ 
is (pre-factors omitted)
\begin{equation*}
\begin{tikzpicture}[baseline=(current bounding box.center)]
    \draw[black, line width=1pt] (0,1) circle (0.4cm);
    
    \draw[black, line width=1pt] (0,-1) circle (0.4cm);
    
    \draw[blue, line width=1pt, decorate, decoration={snake, amplitude=0.5mm, segment length=3mm}]
    (0., 0.6) -- (0., -0.6);  
\end{tikzpicture}
\quad +\frac{1}{2}\quad
\begin{tikzpicture}[baseline=(current bounding box.center)]
    \draw[black, line width=1pt] (0,0.7) to[bend left=45] (1.5,0.7);
    \draw[black, line width=1pt] (0,0.7) to[bend right=45] (1.5,0.7);
    \draw[black, line width=1pt] (0,-0.7) to[bend left=45] (1.5,-0.7);
    \draw[black, line width=1pt] (0,-0.7) to[bend right=45] (1.5,-0.7);
    \draw[blue, line width=1pt, decorate, decoration={snake, amplitude=0.5mm, segment length=3mm}]
    (0., 0.7) -- (0., -0.7);  
    \draw[blue, line width=1pt, decorate, decoration={snake, amplitude=0.5mm, segment length=3mm}]
    (1.5, 0.7) -- (1.5, -0.7);  
\end{tikzpicture}
\quad +\frac{1}{3}\quad
\begin{tikzpicture}[baseline=(current bounding box.center)]
    \draw[black, line width=1pt] (0,0.7) to[bend right=30] (1,0.7); 
    \draw[black, line width=1pt] (1,0.7) to[bend right=30] (2,0.7);
    \draw[black, line width=1pt] (0,0.7) to[bend left=45] (2,0.7); 
    \draw[blue, line width=1pt, decorate, decoration={snake, amplitude=0.5mm, segment length=3mm}]
    (0., 0.7) -- (0., -0.7);  
    \draw[blue, line width=1pt, decorate, decoration={snake, amplitude=0.5mm, segment length=3mm}]
    (1, 0.7) -- (1, -0.7); 
    \draw[blue, line width=1pt, decorate, decoration={snake, amplitude=0.5mm, segment length=3mm}]
    (2, 0.7) -- (2, -0.7);
    \draw[black, line width=1pt] (0,-0.7) to[bend left=30] (1,-0.7); 
    \draw[black, line width=1pt] (1,-0.7) to[bend left=30] (2,-0.7);
    \draw[black, line width=1pt] (0,-0.7) to[bend right=45] (2,-0.7); 
\end{tikzpicture}
\quad +\;\cdots.
\end{equation*}
At order two, there are another two diagrams with bare lines (\text{i.e.} only with the first bare interaction term in the RPA series in Eq. (\ref{eq:RPA_series}) in the main text.) 
\begin{equation*}
\begin{tikzpicture}[baseline=(current bounding box.center)]
\draw[black, line width=1pt] (-0.7, 1+0.7) to[bend right=40] (-0.7, 1-0.7); 
\draw[black, line width=1pt] (-0.342, 1+0.85) to[bend left=30] (0.342, 1+0.85); 
\draw[black, line width=1pt] (-0.342, 1-0.9) to[bend right=30] (0.342, 1-0.9);
\draw[black, line width=1pt] (0.7, 1+0.7) to[bend left=30] (0.7, 1-0.7);
\draw[black, line width=1pt, dashed] (0.7, 1+0.7) to (0.7, 1-0.7);
\draw[black, line width=1pt, dashed] (0.342, 1+0.85) to (0.342, 1-0.9);
\draw[black, line width=1pt, dashed] (-0.7, 1+0.7) to (-0.7, 1-0.7);
\draw[black, line width=1pt, dashed] (-0.342, 1+0.85) to (-0.342, 1-0.9);
\draw[blue, line width=1pt, decorate, decoration={snake, amplitude=0.5mm, segment length=1mm}]
    (0.342, 1-0.9) -- (0.7, 1-0.7);
\draw[blue, line width=1pt, decorate, decoration={snake, amplitude=0.5mm, segment length=1mm}]
    (-0.342, 1-0.9) -- (-0.7, 1-0.7);
\end{tikzpicture}
\quad \longrightarrow \quad
\begin{tikzpicture}[baseline=(current bounding box.center)]
    \draw[black, line width=1pt] (0,0.5) to[bend left=45] (1.5,0.5);
    \draw[black, line width=1pt] (0,0.5) to[bend right=45] (1.5,0.5);
     \draw[black, line width=1pt] (0,-0.8) circle (0.3cm);
     \draw[black, line width=1pt] (1.5,-0.8) circle (0.3cm);     \draw[blue, line width=1pt, decorate, decoration={snake, amplitude=0.5mm, segment length=3mm}]
    (0., 0.5) -- (0., -0.5);  
    \draw[blue, line width=1pt, decorate, decoration={snake, amplitude=0.5mm, segment length=3mm}]
    (1.5, 0.5) -- (1.5, -0.5);  
\end{tikzpicture}
\end{equation*}
Dashed lines correspond to delta functions. There is another spin configuration which amounts to flip the the diagram. It is also interesting to see that the third diagram in the series correspond to a diagram in $(6,)$ with three interactions crossed, while the pre-factors compensate. One can check that all the $26$ connected bare-$U$ diagrams at order three can be found (with some topological identical diagrams grouped) in $(6,)$ with all bare lines, $(4,)$ with one bare line and one second term in Eq. (\ref{eq:RPA_series}), and the third term in the $\frac{1}{2}\mathrm{Tr}\mathrm{ln}[1-W_{G}W_{\Sigma}]$.
\end{document}